\documentclass[%
 aip,
 amsmath,amssymb,
 reprint,%
]{revtex4-1}

\usepackage{graphicx}% Include figure files
\usepackage{dcolumn}% Align table columns on decimal point
\usepackage{bm}% bold math
\usepackage[utf8]{inputenc}
\usepackage[T1]{fontenc}
\usepackage{mathptmx}
\usepackage{amsfonts}
\usepackage{amssymb}
\usepackage{amsmath}
\usepackage{graphicx}
\usepackage{bm}
\usepackage{color}
\usepackage{epsfig}
\usepackage{calc}
\usepackage{ifthen}
\usepackage{subfigure}
\usepackage{graphicx}
\usepackage{epstopdf}
\usepackage{epsfig}

\begin{document}

\preprint{AIP/123-QED}

\title[Sample title]{In-plane terahertz surface plasmon-polaritons coupler based on adiabatic following}
% Force line breaks with \\

\author{Wei Huang}
\altaffiliation{Guangxi Key Laboratory of Optoelectronic Information Processing, Guilin University of Electronic Technology, Guilin 541004, China}%Lines break automatically or can be forced with \\

\author{Xiaowei Qu}
\affiliation{Guangxi Key Laboratory of Optoelectronic Information Processing, Guilin University of Electronic Technology, Guilin 541004, China}

\author{Shan Yin}%
\email{syin@guet.edu.cn}
\affiliation{Guangxi Key Laboratory of Optoelectronic Information Processing, Guilin University of Electronic Technology, Guilin 541004, China}%

\author{Mingrui Yuan}
\affiliation{Center for Terahertz Waves and College of Precision Instrument and Optoelectronics Engineering, Tianjin University, Tianjin 3000072, China}

\author{Wentao Zhang}
\email{zhangwentao@guet.edu.cn}
\affiliation{Guangxi Key Laboratory of Optoelectronic Information Processing, Guilin University of Electronic Technology, Guilin 541004, China}

\author{Jiaguang Han}
\affiliation{Center for Terahertz Waves and College of Precision Instrument and Optoelectronics Engineering, Tianjin University, Tianjin 3000072, China}%

\date{\today}% It is always \today, today,

\begin{abstract}
We propose a robust and broadband integrated terahertz (THz) coupler based on the in-plane surface plasmon polaritons (SPPs) waveguides, conducted with the quantum coherent control -- Stimulated Raman Adiabatic Passage (STIRAP). 
Our coupler consists of two asymmetric specific curved corrugated metallic structures working as the input and output SPPs waveguides, and one straight corrugated metallic structure functioning as the middle SPPs waveguide. From the theoretical and simulated results, we demonstrate that the SPPs can be efficiently transfered from the input to the output waveguides. Our device is robust against the perturbations of geometric parameters, and meanwhile it manifests broadband performance (from 0.3 THz to 0.8 THz) with the high transmission rate over 70$\%$. The in-plane THz coupler can largely simplify the fabrication process, which will make contribution to develop compact and robust integrated THz devices and promote the future applications in all optical network and THz communications.
\end{abstract}

\maketitle

%\begin{quotation}
%The ``lead paragraph'' is encapsulated with the \LaTeX\ 
%\verb+quotation+ environment and is formatted as a single paragraph before the first section heading. 
%(The \verb+quotation+ environment reverts to its usual meaning after the first sectioning command.) 
%Note that numbered references are allowed in the lead paragraph.
%
%he lead paragraph will only be found in an article being prepared for the journal \textit{Chaos}.
%\end{quotation}

\section{Introduction}

Surface plasmon polaritons (SPPs) are electromagnetic waves that propagate along the interface between two materials (a metal and a dielectric) at optical frequency. 
SPPs have attracted increasing attentions due to the abilities of energy confinement and field enhancement \cite{Liu2014, Williams2008,Yin2015}. 
The unique property associated with SPPs has potential applications in superresolution imaging \cite{X.Zhang2008}, high-density optical data storage \cite{C.Genet2007} and sensitivity biosensors \cite{S.Nie1997,H.Yoshida2007,W.L.Barnes2003,C.Huang2007}. 
In the optical regime, SPPs have displayed excellent capabilities in achieving transportation and confinement of photonic energy, which can help to acquire more integrated optical devices.

THz waves, whose vacuum wavelengths range from 30 $\mu m$ to 3 mm, bear outstanding potentials in a number of technological areas, such as medical diagnostics \cite{Fernandez2009}, sensing \cite{B.F.Ferguson2002}, and product quality control and security imaging \cite{K.Kneipp1997}. 
Thus, THz devices are very important for diverse research topics. 
%Due to the wavelength of THz, the length of THz wave device is normally in the scale of $\mu m$ or $mm$. 
In particular, compact THz integrated devices will be the next hotspot. 
%It is straightforward to manufacture THz wave integrated device, with processing THz waves information via SPPs by excited THz waves, due to wavelength of SPPs much smaller than wavelength of THz wave.
Due to the smaller wave vector of the SPPs, the THz coupler based on SPPs waveguide can be scaled down to sub-wavelength region, which is helpful in producting THz integrated devices. 
Since most metals behave as perfect electrical conductor (PEC) at THz frequency, the structureless interface can not bound surface wave efficiently.
In order to realize highly confinement SPPs at THz frequency, corrugated metal structures are proposed to support and propagate the SPPs, called spoof SPPs, which displays similar natures in dispersion properties, subwavelength confinement, field enhancements, and environment sensitivity as those of the optical SPPs \cite{Y.Zhang2018}. %It opens up a new approach for the next generation of various miniaturized terahertz integrated devices.

There are some remarkable papers working on compact THz integrated devices by employing SPPs in previous studies and their devices are made of parallel structured SPPs waveguides \cite{Y.Zhang2018, X.Liu2014}. 
However, those works can hardly realize good permance in broadband since SPPs are strongly dependent on structural parameters \cite{C.Genet2007,Yin2019}. 
In this paper, we propose an asymmetric specific curved corrugated SPPs three-waveguide coupler at THz regime by utilizing quantum coherent control - Stimulated Raman Adiabatic Passage (STIRAP). 
STIRAP is one of the most famous quantum coherent control methods, which is widely used in the quantum optics and quantum systems \cite{Huang20193, Vitanov2017,Vitanov2001}.
The advantages of the STIRAP is that this method can perform high fidelity with robustness against the varying parameters of laser pulse. 
Interestingly, there are many classical systems to obtain robust device against the geometrical parameters and good performance at wide bandwidth by utilizing STIRAP, such as optical waveguide couplers \cite{Huang20191, Longhi2006,Hristova2016}, wireless energy transfer \cite{Rangelov2012} and electronic or SPPs transfer on graphene sheets \cite{Huang20181, Huang20182}. 

Intuitively, we can achieve robust and broadband integrated THz device via STIRAP, which is marvelous for reducing the fabrication cost and enhancing the applicability of THz device, since our design does not require highly precise geometrical parameters and can perform at broad frequency.
Most recently, we proposed a research about robust and broadband integrated THz coupler with multi layers of thin films, by applying the STIRAP in the theoretical prediction \cite{Huang20192}, whose deficiency is the complicated fabrication.
In this paper, we design an in-plane THz coupler based on STIRAP, and demonstrate the robust and broadband performance with the full wave simulation. 
Compared with the previous work, this paper accesses two significance: \textit{i)} we propose brand new layout of producing three corrugated metal waveguides on the same plane and largely simplify the fabrication process; \textit{ii)} we firstly validate the robustness against the geometrical parameters and the broadband versatility of the device via CST full wave simulated results.

The framework of this paper is as followed: starting with the in-plane two parallel metallic waveguides, we determine the structural parameters of the corrugated SPPs waveguides, and then obtain the coupling strength function against the distance between the waveguides in the second section (see Fig. \ref{fig1}). 
Subsequently, instructing by the coupling strength function, we obtain the geometrical parameters of our device based on the three in-plane waveguides(see Fig. \ref{fig2}), by inferring the distances between the input (output) and middle waveguide (shown in Fig. \ref{fig3}). 
In this configuration, our SPPs integrated THz device achieves robustness against geometrical parameters as shown in Fig. \ref{fig4} in section III.
In the next section, we do the full wave simulation with the example (see Fig. \ref{fig5} (a)) and demonstrate that our device can work at wide bandwidth (illustrated in Fig. \ref{fig5} (b)).

\section{Coupling strength function in the parallel waveguides}

In order to obtain the relationship between he coupling strength and the distance between the adjacent waveguides, we start with the configuration of two parallel waveguides. 
Though the parallel waveguides can only work at single band, it can largely simplify the discussion about he coupling strength compared with the curved waveguides.
Note that the SPPs' transfer on curved waveguide is almost identical to straight case \cite{Pandey2013, Cui2013}. 
Therefore, we can ignore the difference of the coupling strength induced by the curved configuration.
Hence the relationship between the coupling strength and gap distance derived in this section can be directly used in the later calculations. 
Indeed, this relationship can be analytically compute by coupled mode theory \cite{Haus1991}. 
However, since it is tough to solve the mode profile of SPPs on the corrugated metallic waveguides using analytical calculations, we firstly investigate two parallel corrugated metallic waveguides to obtain the coupling strength function against the distance between the waveguides by employing numerical simulation software CST. 

\begin{figure}[hbtp]
\centering
\includegraphics[width=0.4\textwidth]{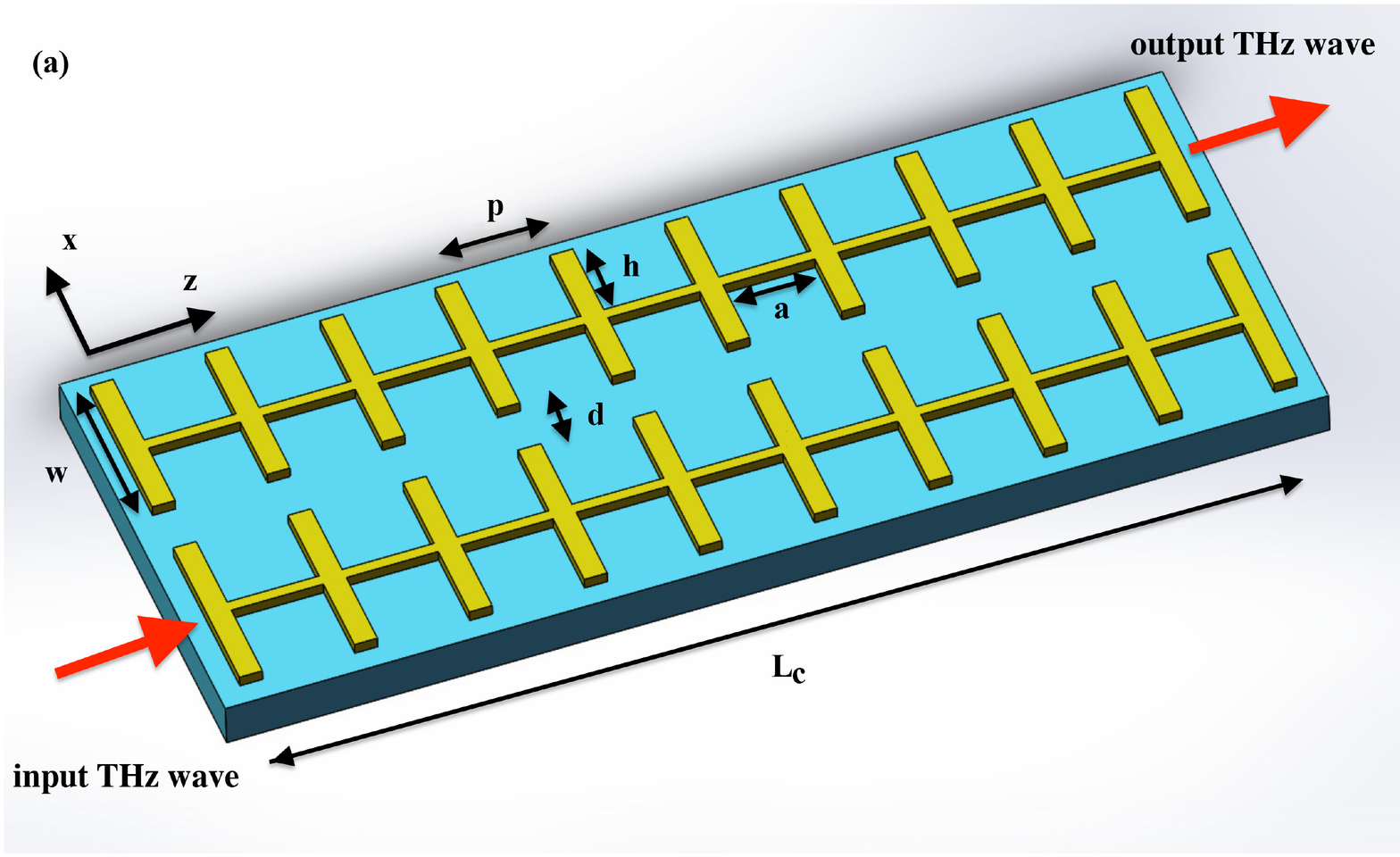}
\includegraphics[width=0.5\textwidth]{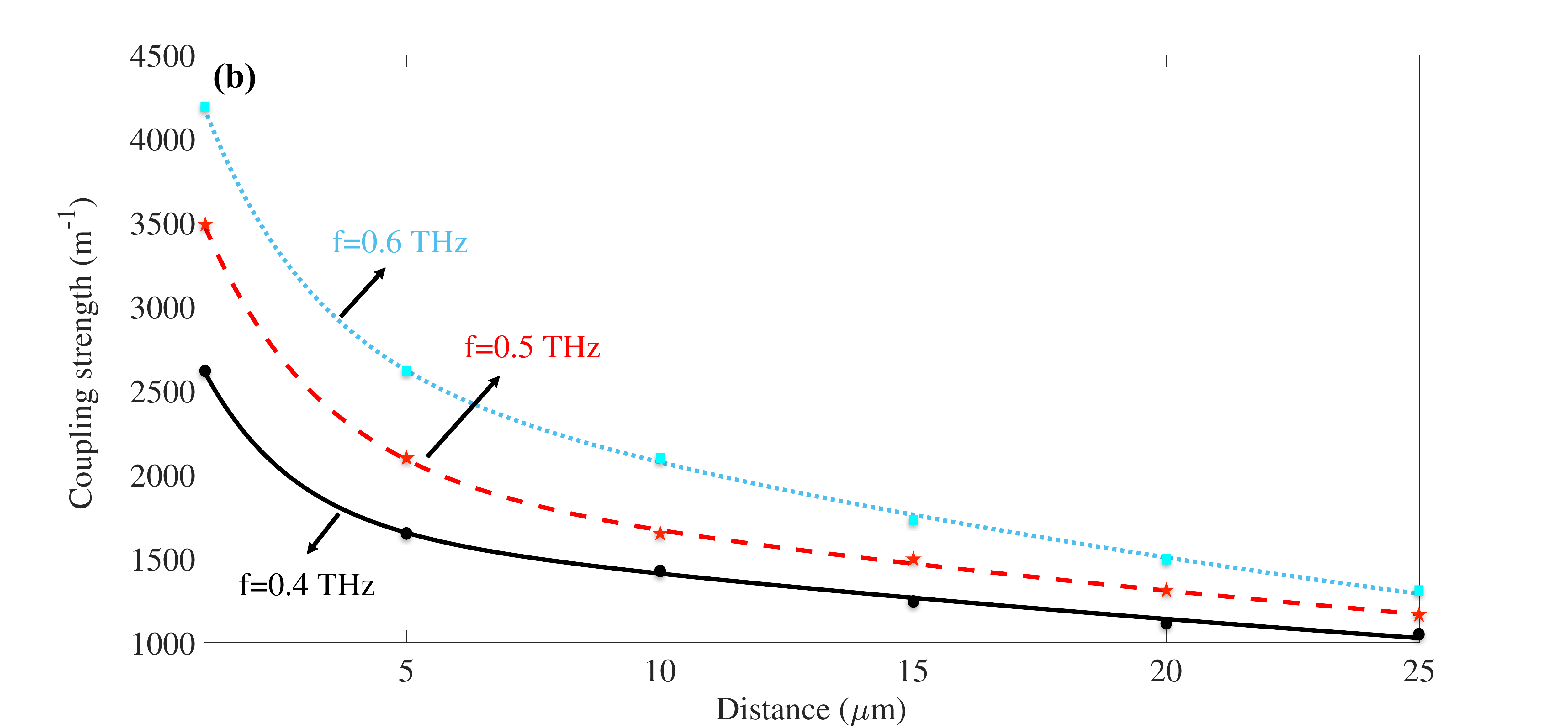}
\caption{ (a) The schematic configuration of the two parallel corrugated metal waveguides, with geometrical parameters width $w$, thickness $t$, depth $h$, grooves width $a$, periodic $p$, distance $d$ and coherent length $L_c$. (b) Coupling strength with the varying distance $d$ at different frequencies. The black circle points, blue star, red square represent coupling strength with 0.4 THz, 0.5 THz and 0.6 THz respectively. The black dot line, blue dashed line and red line are the exponential fitting curves of 0.4 THz, 0.5 THz and 0.6 THz respectively.}
\label{fig1}
\end{figure}

As shown in Fig. \ref{fig1} (a), the diagram displays the schematic model of two parallel corrugated metallic waveguides, where the blue part represents the substrate with a relative dielectric constant of 2.9 and yellow parts represent corrugated metallic waveguides.
The model is composed of two parallel SPPs waveguides with width of $w$ on top of the substrate with thickness of metal $t$, of which two sides are symmetrically corrugated arrays of grooves with depth $h$, width $a$, and periodic $p$. 
The two parallel corrugated metallic waveguides are separated with a distance $d$. 
We set parameters width $w$, thickness $t$, depth $h$,  grooves width $a$, periodic $p$ as 79 $\mu m$, 10 $\mu m$, 40 $\mu m$, 50 $\mu m$, which are consistent with the previous work in Ref. \cite{Y.Zhang2018} to make sure that corrugated thin film waveguide can be easily fabricated.

After determining the above parameters of the corrugated waveguides, we concern the relationship between the coupling strength and the distance between the waveguides. 
In the simulation of parallel waveguides, we can measure the coherent length $L_c$ (the length of SPPs completed transfer from one to another waveguide) from the CST simulation, with different input THz frequencies. 
Subsequently, we can calculate the coupling strength $C$, by using $C = \pi / 2L_c$. 
Fig. \ref{fig1} (b) illustrates the function between coupling strength and distance $d$, with the incident frequencies of 0.4 THz, 0.5 THz and 0.6 THz, which are denoted with the black circle points, blue stars, red squares respectively. 
In addition, the black dot line, blue dashed line and red line are the exponential fitting curves of 0.4 THz, 0.5 THz and 0.6 THz respectively in the Fig. \ref{fig1} (b). 
From the results, we can easily obtain that the coupling coefficient decays exponentially with the increasing distance at a fixed frequency, and the entirety of coupling strength enhances with the increasing incident frequency.
Our simulated results are consistent with coupling mode theory (CMT) and also the same as results in Ref. \cite{Y.Zhang2018}.

\section{Adiabatic following of in-plane SPPs three-waveguide coupler}

\begin{figure}[hbtp]
\centering
\includegraphics[width=0.5\textwidth]{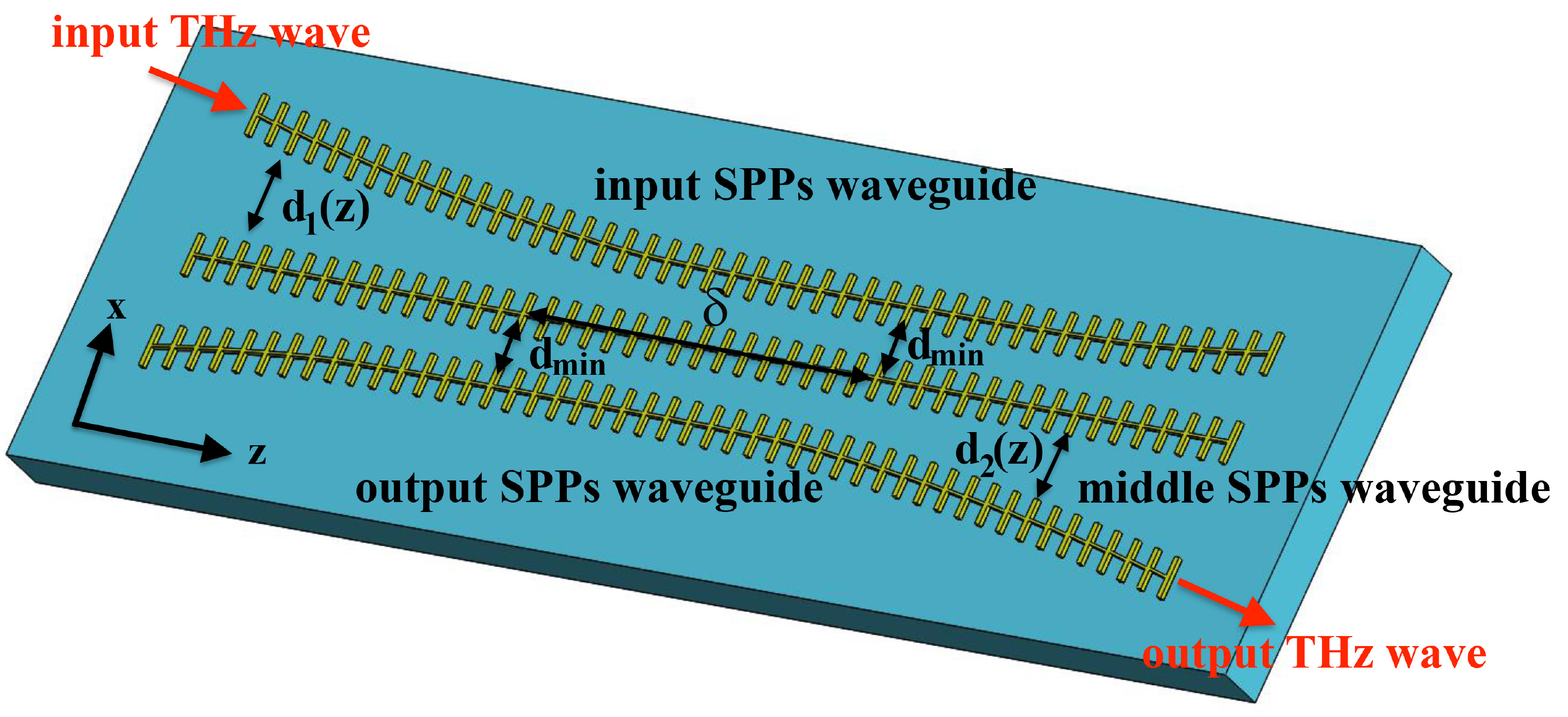}
\caption{The schematic of the asymmetric specific curved corrugated thin films. The distance between input and middle (middle and output) SPPs waveguides is given by the function $d_1(z)$ (and $d_2(z)$), which is derived by coupling strength of STIRAP (Eq. 1). The $d_{min}$ is the minimum distance between input/output SPPs waveguide and middle SPPs waveguide and two minimum distances have the mismatch distance $\delta$. }
\label{fig2}
\end{figure}

In this section, we derive the configuration of the three-waveguide coupler by employing one of famous quantum coherent control - STIRAP. 
The essential element of STIRAP is two Gaussian shaped coupling strengths. 
Thus, the coupling strength between the input/output and middle SPPs waveguide is required as two Gaussian profiles of STIRAP, which can be realized via the well arranged curved corrugated metallic structures. 
In this configuration of the coupling strength, the adiabatic following theory of STIRAP shows that it delivers the complete population (or intensity) transfer from initial state to target state with the robust feature by exploiting anti-intuition sequence \cite{Vitanov2017,Vitanov2001}. 
The coupler consists of two asymmetric specific curved corrugated metallic structures (asinput and output SPPs waveguides) and one straight corrugated metallic structure (as the middle SPPs waveguide).
The distance between input/output and middle SPPs waveguides is given by the function $d_{1/2}(z)$, which is derived based on STIRAP. 
Due to the anti-intuition sequence, there is a mismatching distance $\delta$ between the centers (positions of the minimum distance $d_{min}$) of the input and output waveguides, shown in Fig. \ref{fig2}. 
The previous studies \cite{Longhi2006, Huang20182, Huang20192} had already shown that in the similar configurations,  and the coupling strengths of input (and output) and middle waveguides are obtained as two Gaussian profiles, which can be given by theory of STIRAP. 
The theoretical derivations of applying STIRAP to THz SPPs waveguide coupler have already shown in our previous work \cite{Huang20192}. 
Therefore, we can determine the structural parameters of in-plane SPPs three-waveguide coupler with STIRAP. 
Subsequently, we demonstrate the complete transfer the intensity of SPPs based on ideal case, lossy case. 
At the end of this section, we verify the applicability of our device with robustness against perturbations of geometrical parameters. %and performance at wide frequency range. 

\subsection{Configuration of the in-plane coupler}

Most recently, we have successfully demonstrated the theoretical application of STIRAP to the design of THz SPPs waveguide coupler \cite{Huang20192} and it is the first paper to introduce the coherent quantum control into the device of THz information processing. 
However, the structure in the previous paper is based on the three-layered bent THz SPPs waveguide coupler, which is exhausting to be fabricated the complicated  structure.
To overcome this problem, we propose the in-plane coupling structure with the STIRAP in this paper.

According to the theory of STIRAP, the two coupling strengths are required as two Gaussian shapes with anti-intuition sequence \cite{Vitanov2017,Vitanov2001} to achieve complete SPPs energy transfer. 
Therefore, we require that the coupling strength between the input/output and middle waveguides of our coupler $C_{1/2}$ should be described with Gaussian function, which can be expressed as,

\begin{eqnarray}
C_1(z) = \Omega_0 \exp\left(\dfrac{-\left(z-\delta /2\right)^2}{\tau^2}\right), \notag \\
C_2(z) = \Omega_0 \exp\left(\dfrac{-\left(z+\delta /2\right)^2}{\tau^2}\right);
\end{eqnarray}
where $\delta$ is the mismatching distance and $\tau$ is the full width at half maxima (FWHM) of the Gaussian shapes.
$\Omega_0$ is the maximum coupling strength, which is determined by minimum distance $d_{min}$.

In this example, we choose the Gaussian shapes parameters (in Eq. 1) as $\delta = 600$ $\mu m$, $\tau = 700$ $\mu m$ and $\Omega_0 = 4187 m^{-1}$. 
%Thus, the Gaussian-shaped coupling strengths between input/output and middle SPPs waveguides are given in Fig. \ref{fig3} (a).
There are two reasons for selecting these parameters: firstly, the maximum coupling strength $\Omega_0$ should be corresponding to minimum distance $d_{min}$, and we choose $d_{min} = 1$ $\mu m$, which can be easily fabricated. 
Secondly, two Gaussian shapes require enough overlapping and non-overlapping between each Gaussian shape based on STIRAP. 
The rest parameters of corrugated SPPs waveguides (width $w$, thickness $t$, depth $h$, grooves width $a$, periodic $p$) are the same to those of the parallel waveguides in section II and we do the calculations at 0.6 THz in this example. 

\begin{figure}[hbtp]
\centering
\includegraphics[width=0.5\textwidth]{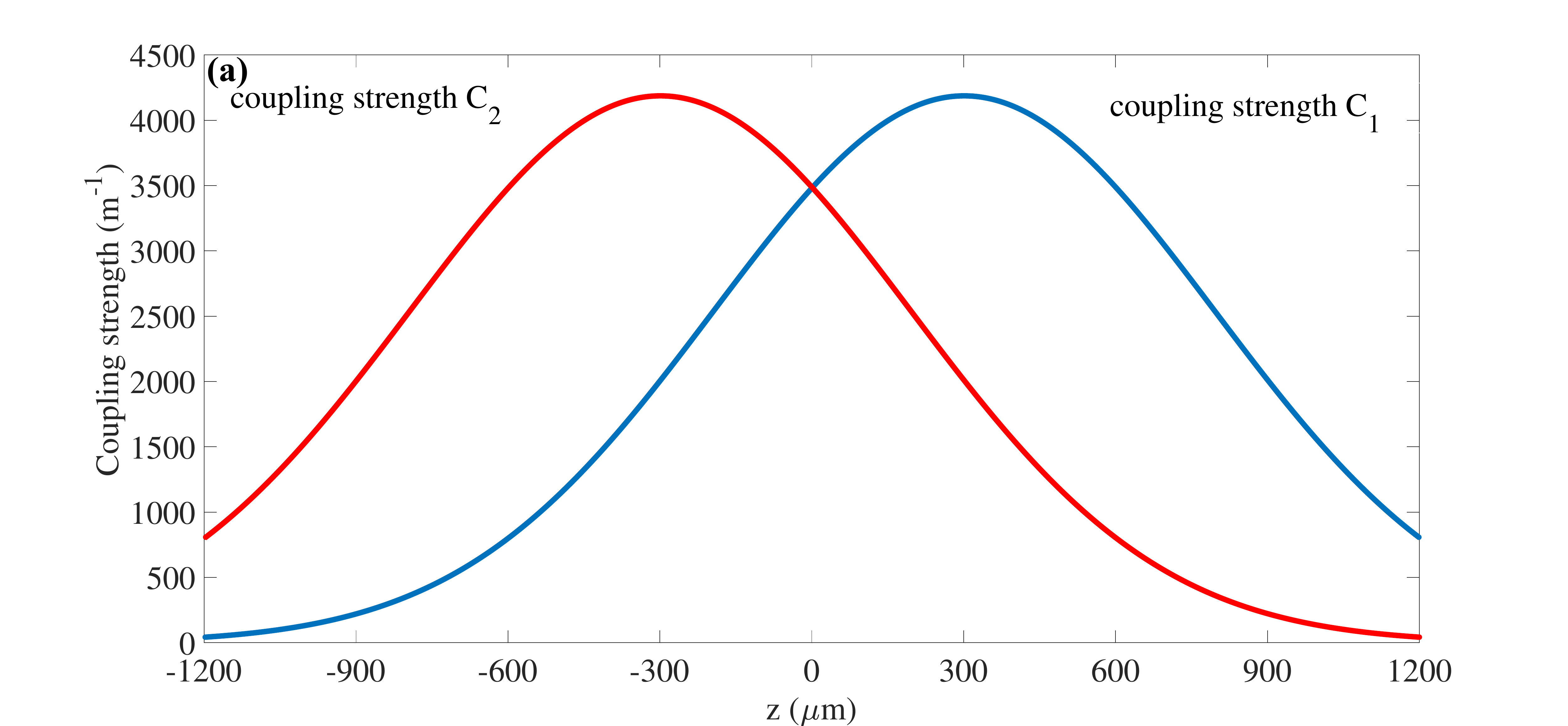}
\includegraphics[width=0.5\textwidth]{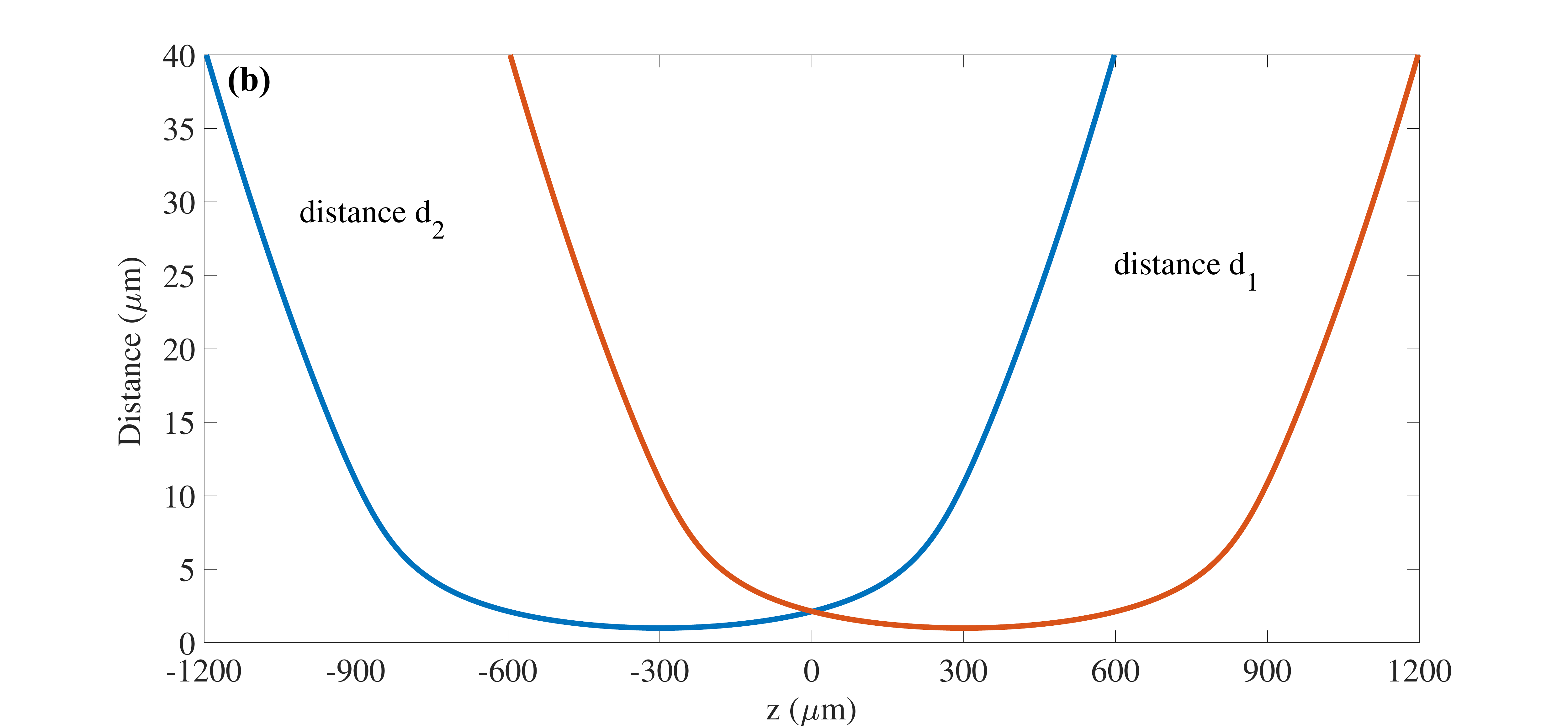}
\includegraphics[width=0.5\textwidth]{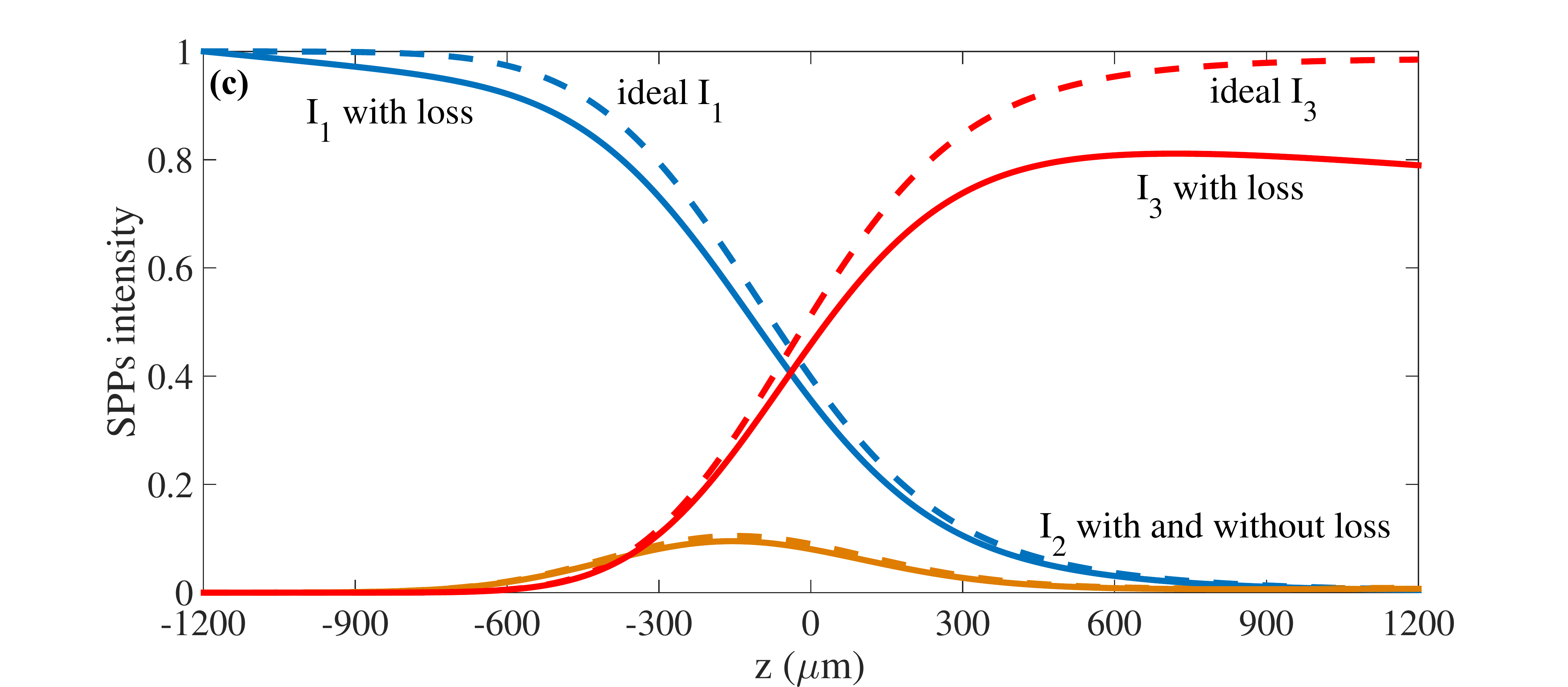}
\caption{Parameter dependent function of the our device at 0.6 THz. We take the geometrical parameters as the $\delta = 600$ $\mu m$ and $d_{min} = 1$ $\mu m$ with the device length from $-1.2$ $mm$ to $1.2$ $mm$. (a) The corresponding coupling strengths of input and middle $C_1$ (output and middle $C_2$) THz SPPs waveguides. (b) The distance between input and middle waveguide $d_1(z)$ (and distance between output and middle waveguide $d_2(z)$). (c) The evolution of SPPs intensities transition of three SPPs waveguides.}
\label{fig3}
\end{figure}

Based on the theory of STIRAP, we calculate the Gaussian-shaped coupling strengths between the input/output and middle SPPs waveguides as shown in Fig. \ref{fig3} (a). 
Associating with the relationship of the coupling strength function against the distance $d$ (see Fig. \ref{fig1} (b)), we can map the Gaussian-shaped coupling strength to the geometrical parameters based on Eq. (1). 
Therefore, we can construct two special-designed curves for input and output SPPs waveguides and straight middle SPPs waveguide, and the distance between input/output and middle waveguide $d_{1/2}(z)$ is given by exponential relationship between coupling strength and $d$. 
The corresponding distances between the input/output and middle waveguides $d_{1/2}(z)$ are obtained in Fig. \ref{fig3} (b). 
Consequently, we can design the geometrical configuration to fix the coupling strength, by given designing by STIRAP.
In this configuration, the evolution of SPPs power transfer is shown in Fig. \ref{fig3} (c). 

From the results of Fig. \ref{fig3} (c), we can achieve complete SPPs power transition from the input to output SPPs waveguides in the ideal case (without lossy), shown in the dashed line in Fig. \ref{fig3} (c).
Furthermore, we consider the lossy case in our calculations with 8 dB/cm loss rate \cite{Y.Zhang2018}, and the SPPs power has the exponential decay along with propagation SPPs' distance. 
In the lossy case, we still can obtain the power transition rate of 80 $\%$ with the solid line in Fig. \ref{fig3} (c). 
Therefore, our coupler can completely transfer the power of SPPs from input to output SPPs waveguide. 

%Based on the theoretical derivations of mimicking the STIRAP to the THz SPPs waveguide coupler \cite{Huang2019}, we design the two circular curves THz SPPs waveguides, with respective input and output SPPs waveguides and radius of circular $R$. Furthermore, we have mismatch $\delta$ between the two center points of two (input/output) SPPs waveguides. The minimum distance between input/output and middle THz SPPs waveguide takes the notation as $d_{min}$. In this example, we set up the parameters as $R = 50$ $mm$, the $\delta = 500$ $\mu m$ and $d_{min} = 1$ $\mu m$. When we have these geometrical parameters, we can obtain the $C_1$ (and $C_2$), with respective to coupling strengths between input (and output) and middle THz Spps waveguides (see Fig. \ref{fig3}(a)). Due to the two geometrical circular curves of input/output SPPs waveguide, we can obtain the two Gaussian shapes of coupling strengths. Therefore, it can deliver the STIRAP into transition of SPPs intensity from input to output waveguide. From the Fig.3(b), we can achieve complete SPPs intensity transition from the input to output SPPs waveguide in the ideal case (without lossy), shown in dashed line. Furthermore, we consider the lossy case in our calculations with 8 dB/cm loss rate \cite{Y.Zhang2018}, which the SPPs intensity has the exponential decay along with propagation SPPs' distance. In the lossy case, we still can obtain the intensity transition rate 80 $\%$ (see Fig. \ref{fig3}(b)).

\subsection{Robustness of our device against the geometrical parameters }
\begin{figure}[hbtp]
\centering
\includegraphics[width=0.5\textwidth]{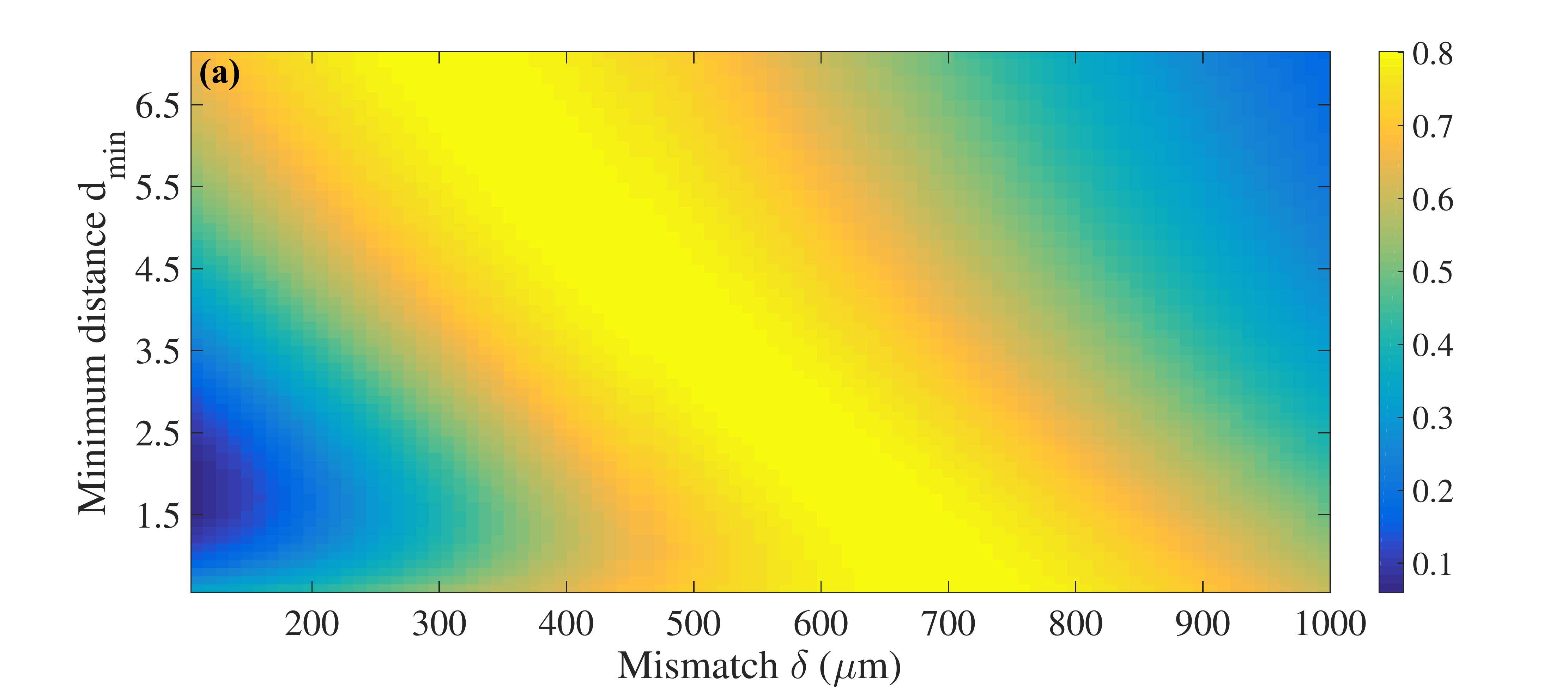}
\includegraphics[width=0.5\textwidth]{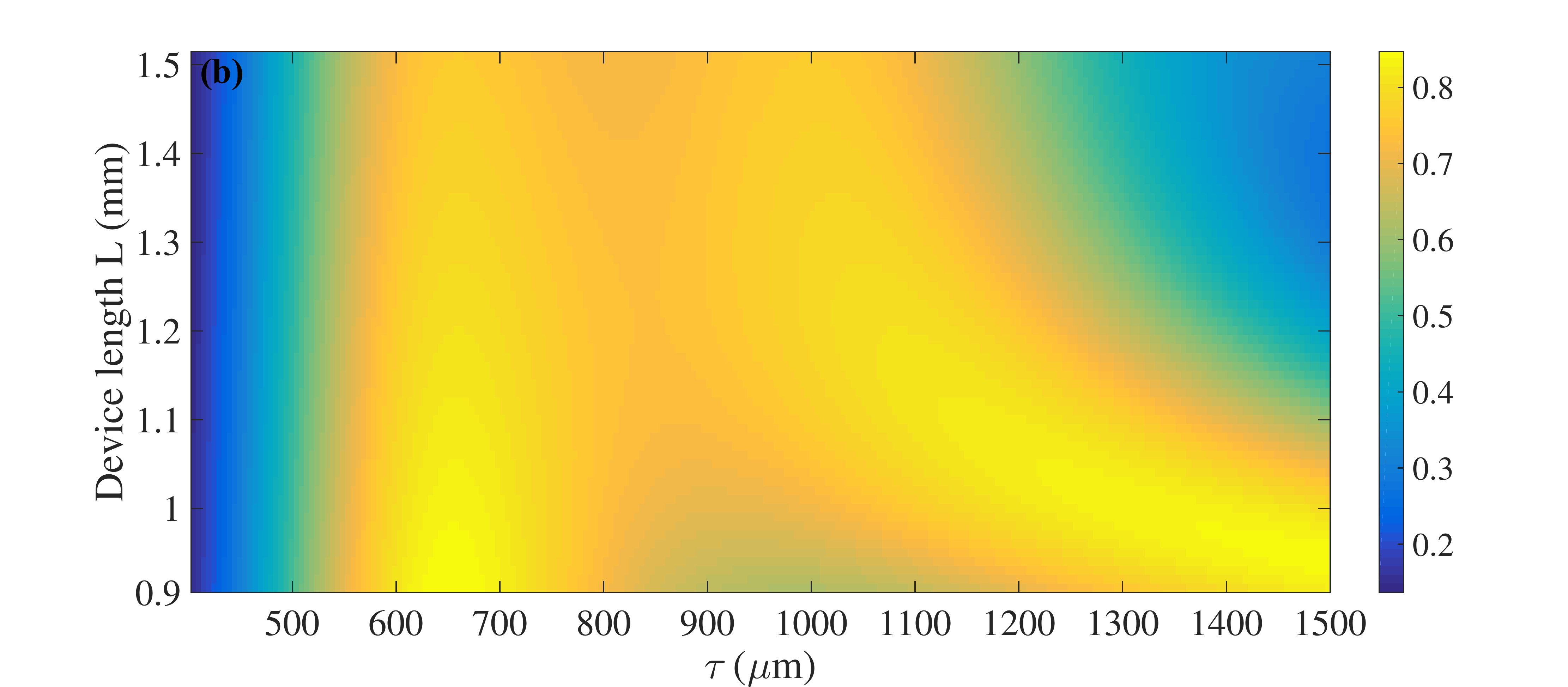}
\caption{Robustness of the geometrical parameters and we do the calculations at 0.6 THz. (a) The final transmission rate of our device with varying the mismatch $\delta$ and minimum distance $d_{min}$, with fixed device length $L = 2400$ $\mu m$ and $\tau = 700$ $\mu m$. (b) The transmission rate of our device with varying device lengths $L$ and $\tau$, with fixing $\d{min}$ = 1 $\mu m$ and mismatch distance $\delta$ = 600 $\mu m$. }
\label{fig4}
\end{figure}

%The reason we introduce the STIRAP into the SPPs waveguide transfer is that the STIRAP is a very robust method in the quantum optics domain \cite{Vitanov2017, Vitanov2001} from the theory of quantum control technology. 
%There are many applications of STIRAP to enhance the robustness of the classical system, such as optical waveguides \cite{Longhi2006}, graphene SPPs transfer \cite{Huang20182} and THz SPPs transfer \cite{Huang20192}.
%STIRAP is a very robust method in the quantum optics domain \cite{Vitanov2017, Vitanov2001} from the theory of STIRAP and we had demonstrated this feature in the previous work \cite{Huang20192}. 
Now we demonstrate the robustness of our device.
We firstly define the final transmission rate as $P_{out}/P_{in}$, where $P_{in/out} = |E_{in/out}|^2$ and $E_{in} (E_{out})$ is the intensity of left-hand-side input (right-hand-side output) SPPs waveguide.
To validate the robustness of our proposed design by varying against geometric parameters, we plot the final transmission rate versus to the offset between two centers of curves $\delta$ (from 100 to 1000 $\mu m$) and the minimum distance $d_{min}$ (from 0.5 to 7 $\mu m$), with the fixed device length $L = 2400$ $\mu m$ and $\tau = 700$ $\mu m$, and results are shown as Fig.\ref{fig4} (a). 
Subsequently, we calculate the final transmission rate against different device lengths $L$ (from $0.9$ $mm$ to $3$ $mm$) and varying $\tau$ (from 400 $\mu m$ to 1500 $\mu m$), by fixing $d_{min}$ = 1 $\mu m$ and mismatch distance $\delta$ = 600 $\mu m$, which is shown in Fig. \ref{fig4} (b). 

It is obviously to see that even though our proposed device has a large perturbation on the geometric parameters (minimum distance $d_{min}$, mismatch distance $\delta$, device length $L$ and $\tau$), the final transmission rate still relatively maintains good performance. 
Therefore, our proposed device is also robust against the varying geometric parameters. 
The advantage of our proposed device is the avoidance of high-precision manufacture process, which can achieve low-cost and high fidelity devices.

\section{Broadband performance of our coupler validated with full wave simulation}
\begin{figure}[hbtp]
\centering
\includegraphics[width=0.5\textwidth]{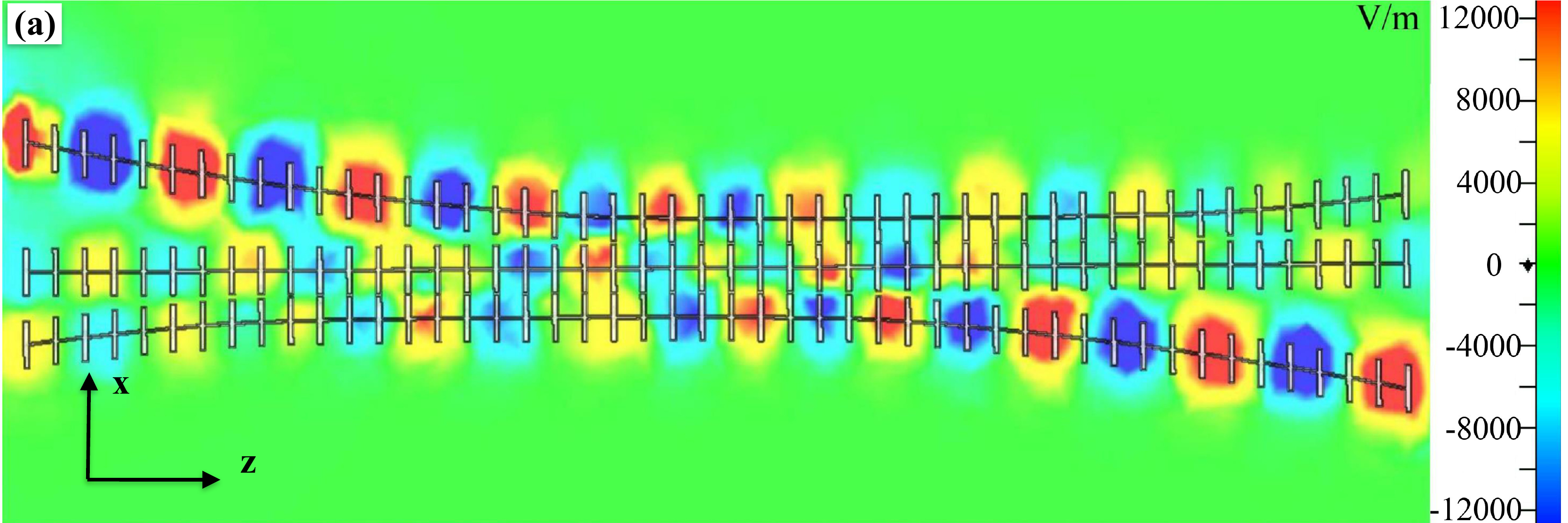}
\includegraphics[width=0.5\textwidth]{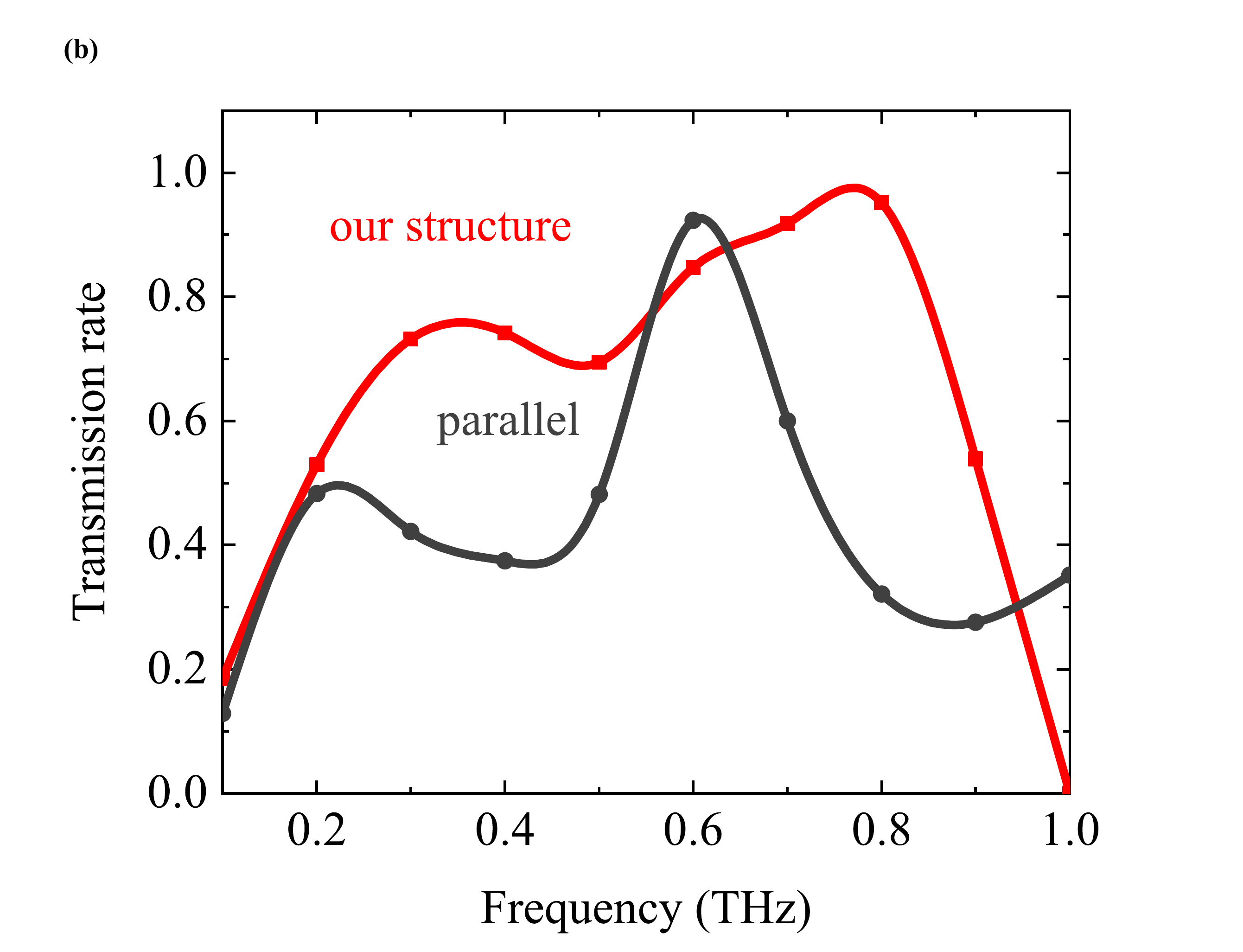}
\caption{Full wave simulated results of our device. (a) The distribution of E-field based on CST software simulation of our device at 0.6 THz. (b) Function of the final transmission rate with different frequencies. The black line shows the transmission rate of the parallel configuration device and red line is the transmission rate of our device. }
\label{fig5}
\end{figure}

In this section, we employ the same corrugated SPPs waveguide and geometrical parameters of our device in section III to do the full wave simulation. 
The simulated results can convincingly validate the theoretical derivation based on STIRAP.
At the beginning, we run the simulation of our device to obtain the field distribution as shown in Fig. \ref{fig5} (a).
Subsequently, we plot the final transmission rate against excited frequency varying from 0.1 to 1 THz with full wave simulation, which can be proved the broadband of our proposed design (see Fig. \ref{fig5} (b)).

We import the SPPs intensity on the left-hand-side input SPPs waveguide and run the simulation in the CST software.
As shown in Fig. \ref{fig5} (a), we can obviously acquire that the majority (nearly 90 $\%$) of SPPs intensity flows to the output of third SPPs waveguide. 
The rest energy (about 10 $\%$) of SPPs leaks to right-hand-side of input/middle SPPs waveguides and left-hand-side of input/middle/output SPPs waveguides, due to the drawing error and reflection effect. 
In spite of few energy leaking to others ports, we still can claim that SPPs intensity complete efficient transfers from the input to output SPPs waveguide.

To visualize the broadband performance of our device, we compare the final transmission rates of our curved three-waveguide device and the parallel waveguides (see Fig. \ref{fig1} (a)) at different frequencies. 
Take the example of the final transmission rates at 0.6 THz. 
Firstly, we run the simulation of parallel configuration and put the monitor port on the second the SPPs waveguide at coherent length $L_c$, which can be calculated by the coupling strength $C = 4187$ $m^{-1}$ at 0.6 THz with gap $g = 1$ $\mu m$ and $L_c = \pi / 2 C$ = 375 $\mu m$. 
In the CST simulation of parallel configuration, we can obtain the intensity integration of monitor port $P_{out}$ with different excited frequency at the coherent length $L_c$.
By varying the input frequencies of THz waves in simulation, we get the final transmission rate $P_{out}/P_{in}$ in the parallel configuration, as shown with the black line in Fig. \ref{fig5} (b).
Subsequently, we change to our device in CST simulation and measure intensity integration of monitor port $P_{out}$ at right-hand-side output SPPs waveguide. 
Thus, we can calculate the final transmission rate by varying the frequencies from 0.1 THz to 1 THz, shown with the red line in Fig. \ref{fig5} (b). 

With the varying input frequencies, the coupling strength function of adjacent SPPs waveguides has large perturbation, due to the mode profile of SPPs waveguide has been modified. 
We all know that the parallel coupling structure is Rabi oscillation model, which is not a robust configuration.
The parallel configuration can not suffer the perturbations of excited frequency (as shown with the black line in Fig. \ref{fig5} (b)).
Consequently, the parallel configuration can only transfer the energy at 0.6 THz and the intensity of SPPs drops rapidly with different input frequencies. 
Oppositely, it is easy to observe that the final transmission rate of our device can operate from 0.3 THz to 0.8 THz with transmission rate higher than 70 $\%$, which is shown with the red line in Fig. \ref{fig5} (b).
Therefore, we can claim that our device have much better performance at bandwidth than the parallel configuration. The advantage of this feature is that our design can largely enhance the universality of THz integrated device, in other words, we can simultaneously transfer information at different frequency channels within one fixed device instead of multiple couplers working at single band.

%There is a remarkable paper whose experiment results showed that the in-plane the structure can support the surface mode of SPPs \textcolor{red}{\cite{}}. Although the experiment was accomplished in the microwave regime, the experiment results were fundamentally consistent with the simulation results under the THz frequency. Therefore, our proposed structure can be also applied in the microwave range in the experiment because both the CSP strips and our design have the same geometric structure sizes.
%In the experiment, the structure can be manufactured using the standard printed circuit broad fabrication process on a three-layer copper-clad laminate. SMA connectors can be employed to provide the input microwave signal, and a section of tapered co-planar waveguide (CPW) is inserted between the SMA connector and the proposed metal structure to match the impedance.

\section{Conclusion}
Based on the stimulated raman adiabatic passage (STIRAP) quantum control technique, we have proposed a novel in-plane asymmetric specific curved corrugated SPPs three-waveguide coupler, in which SPPs can be completely transferred from input waveguide to output waveguide in terahertz (THz) region. 
We demonstrate that our design can realize highly efficient transfer with strong robustness against the perturbations of geometric parameters, and also illustrate that our device has a good performance at broadband excited THz waves. 
This finding will make contribution to develop compact and robust integrated THz devices, which will promote the future applications in all optical network and THz communications.

\section*{Acknowledgements}
This work is acknowledged for funding National Science and Technology Major Project (grant no: 2017ZX02101007-003); National Natural Science Foundation of China (grant no: 61565004; 61965005); the Science and Technology Program of Guangxi Province (grant no: 2018AD19058). W.H. is acknowledged for funding from Guangxi oversea 100 talent project; W.Z. is acknowledged for funding from Guangxi distinguished expert project and J.H. is acknowledged for funding from Guangxi Bagui scholar project.

%\bibliography{aipsamp}% Produces the bibliography via BibTeX.

\section*{References}
\begin{thebibliography}{99}
\bibitem{Liu2014} X. Liu, Y. Feng, K. Chen, B. Zhu, J. Zhao, and T. Jiang, Optics Express \textbf{22}, 20107 (2014).

\bibitem{Williams2008} C. R. Williams, S. R. Andrews, S. A. Maier, A. I. Fernández-Domínguez, L. Martín-Moreno, and F. J. García-Vidal, Nature Photonics \textbf{2}, 175 (2008).

%\bibitem{Shen2008} L. Shen, X. Chen, and T. J. Yang, Optics Express. 16, 3326 (2008).
\bibitem{Yin2015} S. Yin, X. Lu, N. Xu, et al, Scientific reports \textbf{5}, 1 (2015).

\bibitem{X.Zhang2008} X. Zhang, and Z. Liu, Nature Materials \textbf{7}, 435441 (2008).

\bibitem{C.Genet2007} C. Genet, and T. W. Ebbesen, Nature \textbf{445}, 3946 (2007).

\bibitem{S.Nie1997} S. Nie, and S. R. Emory, Science \textbf{275}, 11021106 (1997).

\bibitem{H.Yoshida2007} H. Yoshida, Y. Ogawa, Y. Kawai, S. Hayashi, A. Hayashi, C. Otani, E. Kato, F. Miyamaru, and K. Kawase, Applied Physics Letters \textbf{91}, 253901 (2007).

\bibitem{W.L.Barnes2003} W. L. Barnes, A. Dereux, and T. W. Ebbesen, Nature \textbf{424}, 824830 (2003) 

\bibitem{C.Huang2007} C. Huang, and Y. Zhu, Active and Passive Electronic Components \textbf{2007}, 113 (2007).

\bibitem{Fernandez2009} A. I. Fernandez-Dominguez, E. Moreno, L. Martín-Moreno, et al, Optics letters \textbf{34}, 2063 (2009).

\bibitem{B.F.Ferguson2002}B. F. Ferguson, and X. C. Zhang, Nature Materials \textbf{1}, 26 (2002).

\bibitem{K.Kneipp1997} K. Kneipp, Y. Wang, H. Kneipp, L. T. Perelman, I. Itzkan, R. R. Dasari, and M. S. Feld, Physical Review Letters \textbf{78}, 1667 (1997).

\bibitem{Y.Zhang2018} Y. Zhang, Y. Xu, C. Tian, Q. Xu, X. Zhang, Y. Li, X. Zhang, J. Han, and W. Zhang, Photonics Research \textbf{6}, 18 (2018).

\bibitem{X.Liu2014} X. Liu, Y. Feng, K. Chen, B. Zhu, J. Zhao, and T. Jiang, Optics express \textbf{22}, 20107 (2014).

\bibitem{Yin2019} S Yin, F Hu, X Chen, et al, Journal of Optics \textbf{21}, 025101(2019).

\bibitem{Huang20193} W. Huang, S. Yin, B. Zhu, et al. Physical Review A \textbf{100}, 063430 (2019).

\bibitem{Vitanov2017} N. V. Vitanov, A. A. Rangelov, B. W. Shore, et al, Reviews of Modern Physics \textbf{89}, 015006 (2017).

\bibitem{Vitanov2001} N. V. Vitanov, T. Halfmann, B. W. Shore, et al, Annual review of physical chemistry \textbf{52}, 763 (2001).

\bibitem{Huang20191} W. Huang, L. K. Ang, and E. Kyoseva, Journal of Physics D: Applied Physics \textbf{53}, 035104 (2019).

\bibitem{Longhi2006} S. Longhi S, Physical Review E \textbf{73}, 026607 (2006).

\bibitem{Hristova2016} H. S. Hristova, A. A. Rangelov, G. Montemezzani, et al, Physical Review A \textbf{93}, 033802 (2016).

\bibitem{Rangelov2012}  A. A. Rangelov, and N. V. Vitanov, Annals of Physics \textbf{327}, 2245 (2012).

\bibitem{Huang20181} W. Huang, S. Liang, E. Kyoseva, et al, Semiconductor Science and Technology \textbf{33}, 035014 (2018).

\bibitem{Huang20182} W. Huang, S. Liang, E. Kyoseva, et al, Carbon \textbf{127}, 187 (2018). 

\bibitem{Huang20192} W. Huang, S. Yin, W. Zhang, et al, New Journal of Physics \textbf{21}, 113004 (2019).

\bibitem{Pandey2013} S. Pandey, B. Gupta, A. Nahata, Optics express \textbf{21}, 24422-24430 (2013).

\bibitem{Cui2013} T. J. Cui, X. Shen, Terahertz Science and Technology \textbf{6}, 147 (2013).

\bibitem{Haus1991} H. A. Haus, W. Huang, Proceedings of the IEEE \textbf{79}, 1505 (1991).

\end{thebibliography}

\end{document}